\documentclass[runningheads]{llncs}
\usepackage{graphicx}
\usepackage{amssymb,amsmath} 
\usepackage{hyperref}

\usepackage{pygmentex}

\begin{document}
\title{Voting Theory in the Lean Theorem Prover\thanks{Postprint of the paper in Proceedings of LORI-VIII (Springer) with two typos fixed.}}

\author{Wesley H. Holliday\inst{1}\orcidID{0000-0001-6054-9052} \and
Chase Norman\inst{1}\orcidID{0000-0001-8954-3770} \and
Eric Pacuit\inst{2}\orcidID{0000-0002-0751-9011}}
\authorrunning{Holliday, Norman, and Pacuit}
\institute{University of California, Berkeley  \\
\email{\{wesholliday,c\_\}@berkeley.edu}\\
\and
University of Maryland\\
\email{epacuit@umd.edu}}
\maketitle              % typeset the header of the contribution
\begin{abstract}
There is a long tradition of fruitful interaction between logic and social choice theory. In recent years, much of this interaction has focused on computer-aided methods such as SAT solving and interactive theorem proving. In this paper, we report on the development of a framework for formalizing voting theory in the Lean theorem prover, which we have applied to verify properties of a recently studied voting method. While previous  applications of interactive theorem proving to social choice (using Isabelle/HOL and Mizar) have focused on the verification of impossibility theorems, we aim to cover a variety of results ranging from impossibility theorems to the verification of properties of specific voting methods (e.g., Condorcet consistency, independence of clones, etc.). In order to formalize voting theoretic axioms concerning adding or removing candidates and voters, we work in a variable-election setting whose formalization makes use of dependent types in Lean.

\end{abstract}

\section{Introduction}

There is a long tradition of fruitful interaction between logic and social choice theory.  Both Kenneth Arrow \cite[p.~154]{Arrow2014}  and Amartya Sen \cite[p.~108]{Sen2017} have noted the influence of mathematical logic on their thinking about the foundations of social choice theory.  Early work using logical methods in social choice theory includes  Murakami's  \cite{Murakami1968}  application of  results about three-valued logic to the analysis of voting rules, Rubinstein's \cite{Rubinstein1984}  proof of the equivalence between multi-profile and single-profile approaches to social choice, and Parikh's \cite{Parikh1985} development of a logic of games to study social procedures.  There is now a rich literature developing logical systems that can formalize results in social choice theory (see, e.g., \cite{Pauly2008,AgotnesVanDerHoekWooldridge2009,Nipkow2009,Tang2011,TroquardVanDerHoekWooldridge2011,Endriss2011,GrandiEndriss2013,CinaEndriss2016,PacuitYang2016,HollidayPacuit2020}). 

In recent years, much of the research on logic and social choice has focused on computer-aided methods such as SAT solving and interactive theorem proving \cite{Geist2017}.  The first applications of interactive theorem proving used Isabelle/HOL  \cite{Nipkow2009} and Mizar \cite{Wiedijk2007} to formalize different proofs of  Arrow's Impossibility Theorem \cite{Geanakoplos2005}.  More recently, \cite{Brandt2018b} and \cite{Eberl2019} used Isabelle to verify impossibility theorems from \cite{Brandt2018} and \cite{Brandl2018}, respectively.  These projects demonstrate, as Nipkow \cite{Nipkow2009} notes,  that ``social choice theory turns out to be perfectly suitable for mechanical theorem proving" (p.~303).  In this paper, we provide further evidence of this by developing a framework for formalizing voting theory using an interactive theorem prover. 

But why formalize? One obvious benefit of such a project is the verification of the correctness of mathematical claims in voting theory. Several published claims, including Arrow's \cite{Arrow1951} original statement of his impossibility theorem (for more than 3 candidates), Baigent's \cite{Baigent1987} variation involving ``weak IIA'' (in the case of 3 candidates), and Routley's \cite{Routley1979} claimed generalization of Arrow's theorem to infinite populations, were disproved by counterexamples (see \cite{Blau1957}, \cite{Campbell2000}, and \cite{Blau1979}). Second, formalization allows us to carefully track  which assumptions---e.g., about voter preferences, cardinalities, choice of primitive concepts, etc.---are needed for which results, leading to generalizations and perhaps even new avenues for research. Third, formalization may eventually facilitate automated search of the corpus of proved results for use by researchers in proving new results.

For our formalization project we chose to use the Lean theorem prover \cite{Lean}, a framework that supports both interactive and automated theorem proving.   Lean's kernel is based on dependent type theory and   implements a version of the calculus of inductive constructions \cite{Coquand1988}  and Martin-L\"of type theory \cite{MartinLof1984}. There is an extensive and actively maintained  library of mathematical results formalized in Lean (see \url{https://leanprover-community.github.io/mathlib_docs/}). In addition, Lean is the system chosen for the Formal Abstracts project initiated by Thomas Hales (see \url{https://formalabstracts.github.io}). 
 
Our aim was to use Lean to verify results involving axioms for voting methods (e.g., Condorcet consistency, independence of clones, etc.). In order to formalize axioms concerning adding or removing candidates and voters, we work in a variable-election setting whose formalization makes use of dependent types, as explained in Section \ref{Framework}.  In Section \ref{Theorems}, we discuss our formal verification of results from \cite{HP2020b} about a recently studied voting method, Split Cycle (defined in Example \ref{SplitCycleEx1} below), illustrating the usefulness  of our  framework. We conclude in Section \ref{Conclusion} with directions for further work. All of the code for our project is available at \url{https://github.com/chasenorman/Formalized-Voting}.
 
\section{Framework}\label{Framework}

In this section, we define the basic objects of voting theory: profiles, social choice correspondences, etc. We first give standard set-theoretic definitions and then their type-theoretic counterparts in Lean syntax. After defining these objects, we discuss our formalization of standard axioms used to evaluate voting procedures.

\subsection{Profiles}

For our set-theoretic definitions, we fix infinite sets $\mathcal{V}$ and $\mathcal{X}$ of voters and candidates, respectively. Given $X\subseteq\mathcal{X}$, let $\mathcal{B}(X)$ be the set of all binary relations on $X$. Instead of thinking of a binary relation as a set of ordered pairs, here it is more convenient to think of a binary relation on $X$ as a function $S: X\times X \to \{0,1\}$. In fact, to better match our Lean formalization, we ``curry'' all functions with multiple arguments, transforming them into functions with single arguments that output functions. Thus, we regard a binary relation on $X$ as a function  $S:X\to \{0,1\}^X$, where $\{0,1\}^X$ is the set of functions from $X$ to $\{0,1\}$. For any $x\in X$, $S(x): X\to \{0,1\}$, and $S(x)(y)=1$ means that the binary relation $S$ holds of $(x,y)$. In what follows, we write `$xSy$' instead of $S(x)(y)=1$.

\begin{definition}\label{ProfileDef} \textnormal{For $V\subseteq\mathcal{V}$ and $X\subseteq\mathcal{X}$, a \emph{$(V,X)$-profile} is a map $\mathbf{Q}:V\to \mathcal{B}(X)$. We write `$\mathbf{Q}_i$' for the relation $\mathbf{Q}(i)$. Given a $(V,X)$-profile $\mathbf{Q}$, let $V(\mathbf{Q})$ be $V$ and $X(\mathbf{Q})$  be $X$. We then define a function $\mathsf{Prof}$ that assigns to each pair $(V,X)$ of $V\subseteq\mathcal{V}$ and $X\subseteq\mathcal{X}$ the set $\mathsf{Prof}(V,X)$ of all $(V,X)$-profiles. Finally, define $\mathsf{PROF} = \bigcup_{V\subseteq\mathcal{V},X\subseteq\mathcal{X}}\mathsf{Prof}(V,X)$.}\end{definition}

Depending on the application, one can interpret $x\mathbf{Q}_i y$ to mean either (i) that voter $i$ strictly prefers $x$ to $y$ or (ii) that voter $i$ strictly prefers $x$ to $y$ or is indifferent between $x$ and $y$. We allow either interpretation for the sake of generality, as different voting theorist select different primitives. Under interpretation (i), we use `$\mathbf{P}$' for a profile; under interpretation (ii), we use `$\mathbf{R}$' for a profile.\footnote{\label{VoterNote}Approach (ii) is more general, since it allows one to distinguish between voter $i$ being \textit{indifferent} between $x$ and $y$, defined as $x\mathbf{R}_iy$ and $y\mathbf{R}_ix$, vs. $x$ and $y$ being \textit{noncomparable} for $i$, defined as \textit{neither} $x\mathbf{R}_iy$ \textit{nor} $y\mathbf{R}_ix$. When the distinction between voter indifference and noncomparability is not needed, approach (i) can be simpler.} A profile $\mathbf{Q}$ is said to be \emph{asymmetric} (\emph{transitive}, etc.) if for every $i\in V$, $\mathbf{Q}_i$ is asymmetric (transitive, etc.). Of course, asymmetric profiles only make sense under interpretation (i), whereas under interpretation (ii), profiles should be reflexive.

To translate Definition \ref{ProfileDef} into Lean, we first think of $V$ and $X$ as types, rather than sets, and then represent the function \textsf{Prof} from Definition \ref{ProfileDef} as follows:\footnote{When writing type expressions, arrows associate to the right, so, e.g., the expression  `\texttt{V $\to$ X $\to$ X $\to$ Prop}' stands for \texttt{V $\to$ (X $\to$ (X $\to$ Prop))}.}
\begin{itemize}
\item[] \texttt{\textcolor{blue}{def} \textcolor{orange}{Prof} : \textcolor{blue}{Type} $\to$ \textcolor{blue}{Type} $\to$ \textcolor{blue}{Type} := }\\\texttt{$\textcolor{blue}{\lambda}$ (V X : \textcolor{blue}{Type}), V $\to$ X $\to$ X $\to$ \textcolor{blue}{Prop}}
\end{itemize}
Here \texttt{Prop} is the type of propositions, which in the definition plays the role of $\{0,1\}$ in the treatment of binary relations mentioned above.  The definition states that \texttt{Prof} is a function that given two types, \texttt{V} and \texttt{X}, outputs the type \texttt{V $\to$ X $\to$ X $\to$ Prop}. Because \texttt{X $\to$ X $\to$ Prop} is the type of binary relations on \texttt{X}, an element of the type \texttt{V $\to$ X $\to$ X $\to$ Prop} can be viewed as a $(V,X)$-profile. Thus, we may think of \texttt{Prof V X} as the type of $(V,X)$-profiles.

One of the most important kinds of information to read off from a profile is whether one candidate is majority preferred to another.

\begin{definition}\label{MajPrefDef} \textnormal{Given a profile $\mathbf{P}$ and $x,y\in X(\mathbf{P})$, we say that \textit{$x$ is majority preferred to $y$ in $\mathbf{P}$} if more voters rank $x$ above $y$ than rank $y$ above $x$.}\end{definition}

In Lean, we formalize Definition \ref{MajPrefDef} as follows:

\begin{itemize}
\item[] \texttt{\textcolor{blue}{def} \textcolor{orange}{majority\_preferred}  $\{$V X : \textcolor{blue}{Type}$\}$ :}

\texttt{Prof V X $\to$ X $\to$ X $\to$ \textcolor{blue}{Prop} := \textcolor{blue}{$\lambda$} P x y,}

\texttt{cardinal.mk $\{$v : V // P v x y$\}$ >
cardinal.mk $\{$v : V // P v y x$\}$}
\end{itemize}
Here `\texttt{$\{$V X : \textcolor{blue}{Type}$\}$}' indicates that \texttt{V} and \texttt{X} are implicit arguments\footnote{See \href{https://leanprover.github.io/reference/expressions.html\#implicit-arguments}{Section 3.3} of the Lean documentation on implicit arguments.} to the function \texttt{majority\_preferred} of type \texttt{Type}. Then \texttt{majority\_preferred} takes in explicit arguments of a $(V,X)$-profile and two candidates  and returns the proposition stating that the cardinality of the set of voters who prefer \texttt{x} to \texttt{y} is greater than the cardinality of the set of voters who prefer \texttt{y} to \texttt{x}. Here the `\texttt{//}' notation indicates that we are identifying the subtype of voters with a certain property, and \texttt{cardinal.mk} gives us the cardinality of the subtype.

Voting theorists are often concerned not only with whether one candidate is majority preferred to another but also, if so, by what margin.

\begin{definition}\label{MarginDef} \textnormal{Given a profile $\mathbf{P}$ and $x,y\in X(\mathbf{P})$, the \textit{margin of $x$ over $y$ in $\mathbf{P}$}, denoted $Margin_\mathbf{P}(x,y)$, is $|\{i\in V(\mathbf{P})\mid x\mathbf{P}_iy\}|-|\{i\in V(\mathbf{P})\mid y\mathbf{P}_ix\}|$.}
\end{definition}
\noindent In Lean, Definition \ref{MarginDef} becomes:

\begin{itemize}
\item[] \texttt{\textcolor{blue}{def} \textcolor{orange}{margin} $\{$V X : \textcolor{blue}{Type}$\}$  [fintype V] : Prof V X $\to$ X $\to$ X $\to$ $\mathbb{Z}$} 

\texttt{:=
    \textcolor{blue}{$\lambda$} P x y, $\uparrow$(finset.univ.filter (\textcolor{blue}{$\lambda$} v, P v x y)).card
        -} \\ \texttt{$\uparrow$(finset.univ.filter (\textcolor{blue}{$\lambda$} v, P v y x)).card}
\end{itemize}
Here `\texttt{[fintype V]}' can be understood as an implicit assumption that \texttt{V} is finite,\footnote{See the Lean community page on \href{https://leanprover-community.github.io/theories/sets.html\#finite-types}{Sets and set-like objects}.} which we make so that we can perform the subtraction in the definition of \texttt{margin}. The \texttt{margin} function takes in explicit arguments of a $(V,X)$-profile and two candidates  and returns the margin of the first over the second; in particular, `\texttt{finset.univ.filter (\textcolor{blue}{$\lambda$} v, P v x y)}' is syntax for constructing  the set \\ $\{v\in V(\mathbf{P})\mid x\mathbf{P}_v y\}$, \texttt{.card} takes the cardinality of the set (a natural number), and $\uparrow$ shifts the type from natural number to integer (so we can subtract).

As usual, we can regard the $Margin_\mathbf{P}$ function as an $|X(\mathbf{P})|\times |X(\mathbf{P})|$ matrix. Since $Margin_\mathbf{P}(x,y)=-Margin_\mathbf{P}(y,x)$, the matrix is skew-symmetric. Treating an integer-valued square matrix as a function from a set $X$ to functions from $X$ to $\mathbb{Z}$, the property of skew-symmetry takes in such a function and outputs the proposition stating that the skew-symmetry equation holds for all pairs:
\begin{itemize}
\item[] \texttt{\textcolor{blue}{def} \textcolor{orange}{skew\_symmetric} $\{$X : \textcolor{blue}{Type}$\}$ : (X $\to$ X $\to \mathbb{Z}$) $\to$ Prop :=} 

\texttt{\textcolor{blue}{$\lambda$} M, $\forall$ x y, M x y = - M y x}.
\end{itemize}
Verifying that $Margin_\mathbf{P}$ is skew-symmetric is trivial using Lean's automation:
\begin{itemize}
\item[] \texttt{\textcolor{blue}{lemma} \textcolor{orange}{margin\_skew\_symmetric} $\{$V X : \textcolor{blue}{Type}$\}$ (P : Prof V X)}\\ \texttt{[fintype V] : skew\_symmetric (margin P) :=} 
 
\texttt{\textcolor{blue}{begin}}

\quad \texttt{unfold margin,}

\quad \texttt{obviously,}

\texttt{\textcolor{blue}{end}}
\end{itemize}
The \texttt{unfold} tactic writes \texttt{margin P} in terms of the definition of \texttt{margin} above, allowing the \texttt{obviously} tactic to fill in the details of the proof of skew-symmetry.

Returning to properties of profiles, one of the most important to consider is whether a profile has a so-called Condorcet winner or even a majority winner.

\begin{definition}\label{CondorcetDef} \textnormal{Given a profile $\mathbf{P}$ and $x\in X(\mathbf{P})$, $x$ is a \textit{Condorcet winner in $\mathbf{P}$} if for all $y\in X(\mathbf{P})$ with $y\neq x$, $x$ is majority preferred to $y$ in $\mathbf{P}$. We say that $x$ is a \textit{majority winner in $\mathbf{P}$} if the number of voters who rank $x$ (and only $x$) in first place is greater than the number of voters who do not rank $x$ in first place.}
\end{definition}
In Lean, Definition \ref{CondorcetDef} becomes:

\begin{itemize}
\item[] \texttt{\textcolor{blue}{def} \textcolor{orange}{condorcet\_winner} $\{$V X : \textcolor{blue}{Type}$\}$ (P : Prof V X) (x : X) :} \\ \texttt{Prop := $\forall$ y $\neq$ x, majority\_preferred P x y}
\item[]
\item[] \texttt{\textcolor{blue}{def} \textcolor{orange}{majority\_winner} $\{$V X : \textcolor{blue}{Type}$\}$ (P : Prof V X) (x : X) :} \\
\texttt{ Prop := cardinal.mk $\{$v : V // $\forall$ y $\neq$ x, P v x y$\}$ > cardinal.mk} \\ \texttt{$\{$v : V // $\exists$ y $\neq$ x, P v y x$\}$}
\end{itemize}

As an example of a more involved proof than the one above showing that the margin matrix is skew-symmetric, we present a proof in Lean that a majority winner is also a Condorcet winner. For this we use several basic theorems provided by Mathlib, including one formalizing the fact that a subtype of a type has cardinality less than or equal to that of the type:\footnote{We have changed variable names and replaced `$\#$' with `\texttt{cardinal.mk}'.}
\begin{itemize}
\item[] \texttt{\textcolor{blue}{theorem} \textcolor{orange}{cardinal.mk\_subtype\_mono} $\{$$\alpha$ : Type u$\}$ $\{$$\varphi$ $\psi$ : $\alpha$ $\to$ Prop$\}$} \\
\texttt{(h : $\forall$ x, $\varphi$ x $\to$ $\psi$ x) :} \\
\texttt{cardinal.mk $\{$x // $\varphi$ x$\}$ $\leq$ cardinal.mk $\{$x // $\psi$ x$\}$}
\end{itemize}
We explain the following Lean proof in detail below:
\begin{itemize}
\item[] \texttt{\textcolor{blue}{lemma} \textcolor{orange}{condorcet\_of\_majority\_winnner} $\{$V X : \textcolor{blue}{Type}$\}$ (P : Prof V X)}

\texttt{[fintype V] (x : X) :}

\texttt{majority\_winner P x $\to$ condorcet\_winner P x :=}

\textcolor{blue}{\texttt{begin}}

\item[\texttt{1.}]\quad  \texttt{intros majority z z\_ne\_x,}

\item[\texttt{2.}]\quad  \texttt{\textcolor{blue}{have} imp1 : $\forall$ v, ($\forall$ y $\neq$ x, P v x y) $\to$ P v x z := \textcolor{blue}{by} finish,}
  
\item[\texttt{3.}]\quad  \texttt{refine lt\_of\_lt\_of\_le \_ (cardinal.mk\_subtype\_mono imp1),}
 
\item[\texttt{4.}]\quad  \texttt{\textcolor{blue}{have} imp2 : $\forall$ v, P v z x $\to$ ($\exists$ y $\neq$ x, P v y x) := \textcolor{blue}{by} finish,}
  
 \item[\texttt{5.}]\quad  \texttt{apply lt\_of\_le\_of\_lt (cardinal.mk\_subtype\_mono imp2),}
  
\item[\texttt{6.}]\quad  \texttt{exact majority, }
  
\textcolor{blue}{\texttt{end}}
\end{itemize}
Since the logical form of what we want to prove, \texttt{majority\_winner P x $\to$ condorcet\_winner P x}, is an implication, we use \texttt{intros} on line \texttt{1} to introduce a name \texttt{majority} for a proof of \texttt{majority\_winner P x}. Then since the consequent, \texttt{condorcet\_winner P x}, is a universal claim, \texttt{$\forall$ y $\neq$ x, majority\_preferred  P x~y}, we introduce a name \texttt{z} for an arbitrary candidate and a name \texttt{z\_ne\_x} for a proof of \texttt{z $\neq$ x}. Our goal is now to prove \texttt{majority\_preferred P x z}. 

The first key move on line \texttt{2} is to prove that everyone who ranks \texttt{x} first ranks \texttt{x} above \texttt{z}, which Lean  does automatically using the \texttt{finish} tactic. Since \texttt{imp1} is a proof of a proposition of the form \texttt{($\forall$ v, $\varphi$ v $\to$ $\psi$ v)}, we can apply the Mathlib theorem \texttt{cardinal.mk\_subtype\_mono} to get a proof \texttt{cardinal.mk\_subtype\_mono imp1} that the number of voters who rank \texttt{x} first is less than or equal to the number of voters who rank \texttt{x} above \texttt{z}.

On line \texttt{3}, we use a Mathlib theorem, \texttt{lt\_of\_lt\_of\_le}, which states that \texttt{n}~$<$~\texttt{m}~$\to$ \texttt{m} $\leq$ \texttt{k} $\to$ \texttt{n} $<$ \texttt{k} (recall that implication associates to the right). Take \texttt{n} to be the number of voters who rank \texttt{z} above \texttt{x}, \texttt{m} to be the number who rank \texttt{x} first, and \texttt{k} to be the number  who rank \texttt{x} above \texttt{z}. Thus, our goal is to prove \texttt{n}~$<$~\texttt{k}, and above we proved  \texttt{m} $\leq$ \texttt{k}. Now \texttt{m} $\leq$ \texttt{k} is not the antecedent of \texttt{n}~$<$~\texttt{m}~$\to$~\texttt{m}~$\leq$~\texttt{k}~$\to$~\texttt{n}~$<$~\texttt{k}, but Lean's \texttt{refine} tactic allows us to insert a placeholder $\_$ for the antecedent, so our goal then becomes proving  \texttt{n} $<$ \texttt{m}. 

To prove \texttt{n} $<$ \texttt{m}, the key move on line \texttt{4} is to prove that everyone who ranks \texttt{z} above \texttt{x} does not rank \texttt{x} first, which Lean does automatically using the \texttt{finish} tactic. Then we can apply \texttt{cardinal.mk\_subtype\_mono} to obtain a proof \texttt{cardinal.mk\_subtype\_mono imp2} that the number \texttt{n} of voters who rank \texttt{z} above \texttt{x} is less than or equal to the number---call it \texttt{m}$'$---of voters who do not rank \texttt{x} first. Thus, we have a proof of \texttt{n~$\leq$~m$'$}, so we can apply the implication  \texttt{n~$\leq$~m$'$~$\to$~m$'$~$<$~m~$\to$~n~$<$ m} provided by the Mathlib theorem \texttt{lt\_of\_le\_of\_lt} to obtain a proof of \texttt{m$'$  $<$ m $\to$ n $<$ m}. Then since \texttt{majority} is exactly a proof of the antecedent of \texttt{m$'$  $<$ m $\to$ n $<$ m}, we obtain a proof of our goal \texttt{n $<$ m}.

\subsection{Functions on profiles}

Next we define two standard kinds of functions in voting theory that input profiles. The first, a \textit{social choice correspondence} (SCC), assigns to a profile a set of candidates, considered tied for winning the election. It is common to consider ``domain restrictions'' on the set of profiles for which the SCC is defined \cite{Gaertner2001}. Thus, one may define an SCC as a function $F$ on some set $\mathcal{D}$ of profiles such that for all $\mathbf{Q}\in\mathcal{D}$, we have ${\emptyset\neq F(\mathbf{Q})\subseteq X(\mathbf{Q})}$. However, for our formalization purposes, it is more convenient to use the following equivalent approach.

\begin{definition} \textnormal{For $V\subseteq\mathcal{V}$ and $X\subseteq\mathcal{X}$, a \textit{social choice correspondence for $(V,X)$}, or $(V,X)$-SCC, is a function  $F: \mathsf{Prof}(V,X)\to \wp(X)$. We abuse terminology and call the set $\{\mathbf{Q}\in\mathsf{Prof}(V,X)\mid F(\mathbf{Q})\neq\emptyset \}$ the \textit{domain} of $F$. We say that $F$ satisfies \textit{universal domain} if its domain is $\mathsf{Prof}(V,X)$.}

\textnormal{Let $\mathsf{SCC}$ be a function that assigns to each pair $(V,X)$ of $V\subseteq\mathcal{V}$ and $X\subseteq\mathcal{X}$ the set of all $(V,X)$-SCCs.}
\end{definition}

We represent the function $\mathsf{SCC}$ in Lean as follows, where \texttt{set X} is the type of subsets of \texttt{X}:\footnote{When writing type expressions, function application binds more strongly than arrow, so `\texttt{Prof V X $\to$ set X}' stands for \texttt{(Prof V X) $\to$ set X}.}
\begin{itemize}
\item[] \texttt{\textcolor{blue}{def} \textcolor{orange}{SCC} := \textcolor{blue}{$\lambda$} (V X : \textcolor{blue}{Type}), Prof V X $\to$ set X}
\end{itemize}
The definition states that \texttt{SCC} is a function that given two types, \texttt{V} and \texttt{X}, outputs the type \texttt{Prof V X $\to$ set X}, which is the type of $(V,X)$-SCCs. 

We formalize universal domain as follows:
\begin{itemize}
\item[] \texttt{\textcolor{blue}{def} \textcolor{orange}{universal\_domain\_SCC} $\{$V X : \textcolor{blue}{Type}$\}$ (F : SCC V X) : Prop :=} \\
\texttt{$\forall$ P : Prof V X, F P $\neq$ $\emptyset$}
\end{itemize}

\begin{example}\label{CondorcetEx1} For any $V$, $X$, consider the Condorcet SCC for $(V,X)$ defined by:
\[\mathrm{Cond}_{(V,X)}(\mathbf{P})=\begin{cases} \{x\} & \mbox{if there is a Condorcet winner $x$ in $\mathbf{P}$} \\ X(\mathbf{P}) & \mbox{otherwise}\end{cases}.\]
The definition states that given a $(V,X)$-profile $\mathbf{P}$, if there is a Condorcet winner---in which case it is unique---then output the set containing the Condorcet winner, and otherwise output all candidates in $X$.

We represent this $(V,X)$-SCC in Lean as follows:
\begin{itemize}
\item[] \texttt{\textcolor{blue}{def} \textcolor{orange}{condorcet\_SCC} $\{$V X : \textcolor{blue}{Type}$\}$ : SCC V X := $\textcolor{blue}{\lambda}$ P, }

\texttt{$\{$x : X |  condorcet\_winner P x $\vee$ $\neg$ $\exists$ y, condorcet\_winner P y$\}$}
\end{itemize}
\end{example}

Most voting methods (e.g., Plurality, Borda, Instant Runoff) are defined not only for a fixed set of voters and candidates but for any set of voters and candidates, which motivates the following definition.

\begin{definition}\label{VSCC} \textnormal{A \textit{variable-election social choice correspondence} (VSCC) is a function $F$ that assigns to each pair $(V,X)$ of a $V\subseteq \mathcal{V}$ and $X\subseteq\mathcal{X}$ a $(V,X)$-SCC. We abuse terminology and call the set $\{\mathbf{Q}\in\mathsf{PROF}\mid F(V(\mathbf{Q}),X(\mathbf{Q}))(\mathbf{Q})\neq\emptyset \}$ the \textit{domain} of  $F$. We say that $F$ satisfies  (\textit{finite}) \textit{universal domain} if the domain of $F$ includes $\{\mathbf{P}\in \mathsf{PROF}\mid V(\mathbf{P}) \mbox{ and }X(\mathbf{P}) \mbox{ nonempty and finite}\}$.}\footnote{Of course, one could also consider the stronger condition that the domain of $F$ contains all profiles even with infinite sets of voters and/or candidates.}
\end{definition}

\noindent An equivalent but perhaps more intuitive approach would define a VSCC to be a function on $\mathsf{PROF}$ (rather than $\wp(\mathcal{V})\times\wp(\mathcal{X})$) such that for each $\mathbf{Q}\in\mathsf{PROF}$, we have $F(\mathbf{Q})\subseteq X(\mathbf{Q})$;\footnote{This is the definition of a \textit{voting method} used in \cite{HP2020b} with the additional stipulations that $F(\mathbf{Q})\neq\emptyset$ and  that $V(\mathbf{Q})$ and $X(\mathbf{Q})$ are nonempty and finite.} abusing terminology, we could then call the set $\{\mathbf{Q}\in\mathsf{Prop}\mid F(\mathbf{Q})\neq\emptyset\}$ the \emph{domain} of the VSCC. However, we have presented Definition \ref{VSCC} above because it nicely connects with our formalization in Lean.

In Lean, we define the type of VSCCs as a \textit{dependent function type}:
\begin{itemize}
\item[] \texttt{\textcolor{blue}{def} \textcolor{orange}{VSCC} : \textcolor{blue}{Type 1} := $\Pi$ (V X : \textcolor{blue}{Type}), SCC V X}
\end{itemize}
Given \texttt{$\alpha$ : Type 1} and \texttt{$\beta$ : $\alpha$ $\to$ $\alpha$ $\to$ Type}, the type \texttt{$\Pi$ y z : $\alpha$, $\beta$ y z} is the type of functions \texttt{f} such that for each \texttt{a b : $\alpha$}, we have that \texttt{f a b} is an element of \texttt{$\beta$~a~b}. In the definition of \texttt{VSCC} above, $\alpha$ is \texttt{Type} and $\beta$ is \texttt{SCC}.  Thus, the definition states that an element of the type \texttt{VSCC} is a function that for any types \texttt{V} and \texttt{X} returns a function of the type \texttt{SCC V X}, i.e., a $(V,X)$-SCC. 

We formalize (finite) universal domain as follows:
\begin{itemize}
\item[] \texttt{\textcolor{blue}{def} \textcolor{orange}{finite\_universal\_domain\_VSCC} (F : VSCC) : Prop :=}\\
    \texttt{$\forall$ V X [inhabited V] [inhabited X] [fintype V] [fintype X],}
    
    \texttt{universal\_domain\_SCC (F V X)}
\end{itemize}

\begin{example} We define the Condorcet VSCC as follows, taking advantage of our definition for any $V$ and $X$ of the Condorcet $(V,X)$-SCC in Example \ref{CondorcetEx1}:
\begin{itemize}
\item[] \texttt{\textcolor{blue}{def} \textcolor{orange}{condorcet\_VSCC} : VSCC := \textcolor{blue}{$\lambda$} V X, condorcet\_SCC}
\end{itemize}
\end{example}
Similarly, we may define VSCCs for Plurality voting, Borda, Instant Runoff, etc. 

The second type of function we consider assigns to a given profile a binary relation on the set of candidates in the profile.

\begin{definition}\textnormal{For $V\subseteq\mathcal{V}$ and $X\subseteq\mathcal{X}$, a \textit{collective choice rule for $(V,X)$}, or $(V,X)$-CCR, is a function  $f: \mathsf{Prof}(V,X)\to \mathcal{B}(X)$. Let $\mathsf{CCR}$ be a function that assigns to each pair $(V,X)$ of $V\subseteq\mathcal{V}$ and $X\subseteq\mathcal{X}$ the set of all $(V,X)$-CCRs.}\end{definition}

\noindent Depending on the application, one can interpret the binary relation $f(\mathbf{Q})$ in one of two ways: $xf(\mathbf{Q})y$ can mean (a) $x$ is strictly preferred to $y$ socially or (b) $x$ is strictly preferred to or tied with $y$ socially.\footnote{As in Footnote \ref{VoterNote}, approach (b) is more general, since it allows one to distinguish between ``social indifference'' and ``social noncomparability'' (for examples of theorems in social choice in which this distinction matters, see \cite{HK2020}). When notions of social indifference and noncomparability are not needed, approach (a) can be simpler.} Once again, there is also the issue of ``domain restrictions.'' Under approach $(a)$, we can mark that the CCR is ``undefined'' on a profile $\mathbf{Q}$ by setting $f(\mathbf{Q})= X(\mathbf{Q})\times X(\mathbf{Q})$. Then we can abuse terminology and call $\{\mathbf{Q}\in\mathsf{Prof}(V,X)\mid f(\mathbf{Q})\neq X(\mathbf{Q})\times X(\mathbf{Q}) \}$ the domain of $f$. Under approach (b), we can mark that the CCR is ``undefined'' on $\mathbf{Q}$ by setting $f(\mathbf{Q})=\emptyset$. Then we can abuse terminology and call $\{\mathbf{Q}\in\mathsf{Prop}(V,X)\mid f(\mathbf{Q})\neq \emptyset\}$ the domain of~$f$.  A CCR $f$ is said to be \textit{asymmetric} (resp.~\textit{transitive}, etc.), if for all $\mathbf{Q}$ in the domain of $f$, $f(\mathbf{Q})$ is asymmetric (transitive, etc.). Of course, asymmetric CCRs only make sense under interpretation (a) above, whereas under interpretation (b), CCRs should be reflexive.

In Lean, our representation of the function $\mathsf{CCR}$ is similar to that of $\mathsf{SCC}$:
\begin{itemize}
\item[] \texttt{\textcolor{blue}{def} \textcolor{orange}{CCR} := \textcolor{blue}{$\lambda$} (V X : \textcolor{blue}{Type}), Prof V X $\to$ X $\to$ X $\to$ \textcolor{blue}{Prop}}
\end{itemize}

\begin{example}\label{SplitCycleEx1} As an example of a CCR, we consider the Split Cycle CCR studied in \cite{HP2020}. The output of the Split Cycle CCR is an asymmetric relation understood as a relation of ``defeat'' between candidates. A candidate $x$ defeats a candidate $y$ in $\mathbf{P}$ just in case the margin of $x$ over $y$ is (i) positive and (ii) greater than the weakest margin in each majority cycle containing $x$ and $y$. To formalize this definition, we first need a definition of a cycle in a binary relation:
\begin{itemize}
\item[] \texttt{\textcolor{blue}{def} \textcolor{orange}{cycle} $\{$X : \textcolor{blue}{Type}$\}$ := \textcolor{blue}{$\lambda$} (R : X $\to$ X $\to$ \textcolor{blue}{Prop}) (c : list X), \\
$\exists$ (e : c $\neq$ list.nil), list.chain R (c.last e) c}
\end{itemize}
Here the function \texttt{cycle} takes in a binary relation \texttt{R} and a list \texttt{c} of elements of \texttt{X} and outputs the proposition stating that (i) there is a proof \texttt{e} that \texttt{c} is not the empty list, and (ii) \texttt{c} is a cycle in \texttt{R}. To express (ii), we use the construction \texttt{list.chain R a c}, where \texttt{R} is a binary relation, \texttt{a} is an element of \texttt{X}, and \texttt{c} is a list of elements of \texttt{X}, which means that \texttt{a} is \texttt{R}-related to the first element of \texttt{c} and that every element in the list \texttt{c} is related to the next element in \texttt{c}. Thus, if we take \texttt{a} as the last element of \texttt{c}, this implies that \texttt{c} is a cycle. Applying \texttt{c.last} to the proof \texttt{e} that \texttt{c} is not the empty list outputs the last element of \texttt{c}.

Now we are ready to define the Split Cycle $(V,X)$-CCR in Lean:
\begin{itemize}
\item[] \texttt{\textcolor{blue}{def} \textcolor{orange}{split\_cycle\_CCR} $\{$V X : \textcolor{blue}{Type}$\}$ : CCR V X :=} 

    \texttt{\textcolor{blue}{$\lambda$} (P : Prof V X) (x y : X), $\forall$ [n : fintype V],}
    
    \texttt{0 $<$ @margin V X n P x y $\wedge$}
    
    \texttt{$\neg$ ($\exists$ (c : list X), x $\in$ c $\wedge$ y $\in$ c $\wedge$}\\
    \texttt{cycle (\textcolor{blue}{$\lambda$} a b, @margin V X n P x y $\leq$ @margin V X n P a b) c)}
\end{itemize}
Recall that the \texttt{margin} function takes as implicit arguments the set \texttt{V} of voters, the set \texttt{X} of candidates, and a proof that \texttt{V} is finite. The \texttt{@} symbol is used when explicitly supplying these implicit arguments. Thus, the definition states that given a profile \texttt{P} and two candidates \texttt{x} and \texttt{y}, the binary relation output by \texttt{split\_cycle\_CCR} holds of \texttt{x}, \texttt{y} if for any proof \texttt{n} that \texttt{V} is finite, the margin of \texttt{x} over \texttt{y} in \texttt{P} (supplying the \texttt{margin} function with \texttt{V}, \texttt{X}, and \texttt{n}) is greater than 0 and there is no list \texttt{c} of elements containing \texttt{x} and \texttt{y} such that \texttt{c} is a majority cycle for which the margin of \texttt{x} over \texttt{y} is less than or equal to every margin in the cycle, i.e., \texttt{c} is a cycle in the binary relation \texttt{R} that holds of \texttt{a}, \texttt{b} just in case the margin of \texttt{x} over \texttt{y} is less than or equal to the margin of \texttt{a} over \texttt{b}.

\end{example}

Once again, we can consider functions that are not restricted to a fixed set of voters and candidates.

\begin{definition}\label{VCCR} \textnormal{A \emph{variable-election collective choice rule} (VCCR) is a function that assigns to each pair $(V,X)$ of a $V\subseteq\mathcal{V}$ and $X\subseteq\mathcal{X}$ a $(V,X)$-CCR.}
\end{definition}

\noindent An equivalent but perhaps more intuitive definition takes a VCCR to be a function $f$ on $\mathsf{PROF}$ (instead of $\mathcal{V}\times\mathcal{X}$) such that for all $\mathbf{Q}\in\mathsf{PROF}$, $f(\mathbf{Q})$ is a binary relation on $X(\mathbf{Q})$.\footnote{This is the definition of a VCCR used in \cite{HP2020} with the aditional stipulation that $V(\mathbf{Q})$ and $X(\mathbf{Q})$ are nonempty and finite.} However, we have presented Definition \ref{VCCR} above because it nicely connects with our formalization in Lean, which as in the case of VSCCs defines the type of VCCRs to be a dependent function type:
\begin{itemize}
\item[] \texttt{\textcolor{blue}{def} \textcolor{orange}{VCCR} := $\Pi$ (V X : \textcolor{blue}{Type}), CCR V X}
\end{itemize}
The definition states that an element of the type \texttt{VCCR} is a function that for any types \texttt{V} and \texttt{X} returns a function of the type \texttt{CCR V X}, i.e., a $(V,X)$-CCR.

\begin{example}\label{SCVCCR} We define the Split Cycle VCCR as follows, taking advantage of our definition for any $V$ and $X$ of the Split Cycle $(V,X)$-SCC in Example \ref{SplitCycleEx1}:
\begin{itemize}
\item[] \texttt{\textcolor{blue}{def} \textcolor{orange}{split\_cycle\_VCCR} : VCCR := \textcolor{blue}{$\lambda$} V X, split\_cycle\_CCR}
\end{itemize}
\end{example}

Any VCCR, regarded as outputting for a given profile (for a given $V$, $X$) a relation of strict social preference or ``defeat,'' can be transformed into a VSCC by assigning to a given profile the set of candidates who are not defeated.\footnote{An alternative approach, also easily formalizable, assigns to a given profile the set of candidates who are weakly socially preferred to all other candidates.}

\begin{definition}\label{InducedDef} \textnormal{Given an asymmetric VCCR $f$, we define the \textit{maximal-element induced} VSCC $f_{M}$ such that for any $V\subseteq\mathcal{V}$, $X\subseteq\mathcal{X}$, and $(V,X)$-profile $\mathbf{P}$, \[f_{M}(V,X)(\mathbf{P})=\{x\in X(\mathbf{P})\mid \forall y\in X(\mathbf{P}),\, (y,x)\not\in f(V,X)(\mathbf{P})\}.\]}
\end{definition}
In Lean, we formalize Definition \ref{InducedDef} as follows:
\begin{itemize}
\item[] \texttt{\textcolor{blue}{def} \textcolor{orange}{max\_el\_VSCC} : VCCR $\to$ VSCC := \textcolor{blue}{$\lambda$} f V X P,} \\
\texttt{$\{$x : X | $\forall$ y : X, $\neg$ f V X P y x$\}$}
\end{itemize}

\begin{example} The Split Cycle voting method \cite{HP2020b} is the maximal-element induced VSCC from the Split Cycle VCCR defined in Example \ref{SCVCCR}: 
\begin{itemize}
\item[] \texttt{\textcolor{blue}{def} \textcolor{orange}{split\_cycle} : VSCC := max\_el\_VSCC split\_cycle\_VCCR}
\end{itemize}
\end{example}

As is well known, any acyclic VCCR (i.e., VCCR that assigns an acyclic CCR to each $V,X$) induces a VSCC satisfying (finite) universal domain:
\begin{itemize}
\item[] \texttt{\textcolor{blue}{def} \textcolor{orange}{acyclic} $\{$X : \textcolor{blue}{Type}$\}$ : (X $\to$ X $\to$ \textcolor{blue}{Prop}) $\to$ \textcolor{blue}{Prop} :=}\\
\texttt{\textcolor{blue}{$\lambda$} Q, $\forall$ (c : list X), $\neg$ cycle Q c}
\item[]
\item[] \texttt{\textcolor{blue}{theorem} \textcolor{orange}{max\_el\_VSCC\_universal\_domain} (f : VCCR)}\\
\texttt{(a : $\forall$ V X [inhabited V] [inhabited X] [fintype V] [fintype X]} \\
\texttt{(P : Prof V X), acyclic (F V X P)) :}\\
\texttt{finite\_universal\_domain\_VSCC (max\_el\_VSCC f) := \dots}
\end{itemize}
The proof can be found in our online repository.

\subsection{Voting axioms}\label{VotingAxioms}

After formalizing the basic objects of voting theory, we  formalized a number of standard axioms by which voting procedures are evaluated (and then proved that the axioms are satisfied by Split Cycle, as explained in Section \ref{Theorems}):
\begin{itemize}
\item[] \textbf{Domain axioms}:
\begin{itemize}
\item \textit{universal domain} (resp.~\textit{finite universal domain}): all profiles (resp.~finite profiles) are in the domain of the VSCC.\\
\end{itemize}
\item[] \textbf{Intra-profile axioms}:
\begin{itemize}
\item \textit{Condorcet criterion}: if there is a Condorcet winner in a profile, that candidate is the unique winner.
\item \textit{Condorcet loser criterion}: if there is a Condorcet loser in a profile---a candidate who loses to every other candidate in a head-to-head majority comparison---that candidate does not win.
\item \textit{Pareto}: if all voters rank candidate $x$ above candidate $y$ in a profile, then $y$ does not win.\\
\end{itemize}
\item[] \textbf{Inter-profile axioms}:
\begin{itemize}
\item \textit{monotonicity}: if $x$ wins in a profile $\mathbf{P}$, and $\mathbf{P}'$ is obtained from $\mathbf{P}$ by voters moving $x$ up in their rankings, then $x$ still wins in $\mathbf{P}'$.
\item \textit{reversal symmetry}: if $x$ is the unique winner in a profile $\mathbf{P}$, then $x$ is not a winner in the profile $\mathbf{P}^r$ obtained from $\mathbf{P}$ by reversing all voters' rankings.\\
\end{itemize}
\item[] \textbf{Variable-voter inter-profile axioms}:
\begin{itemize}
\item \textit{positive involvement}: if $x$ wins in a profile $\mathbf{P}$, and $\mathbf{P}'$ is obtained from $\mathbf{P}$ by adding a new voter who ranks $x$ as their unique first choice, then $x$ still wins in $\mathbf{P}'$.
\item \textit{negative involvement}: if $x$ does not win in a profile $\mathbf{P}$, and $\mathbf{P}'$ is obtained from $\mathbf{P}$ by adding a new voter who ranks $x$ as their unique last choice, then $x$ still does not win in $\mathbf{P}'$.\\
\end{itemize}
\item[] \textbf{Variable-candidate inter-profile axioms}:
\begin{itemize}
\item \textit{strong stability for winners}: if $x$ wins in a profile $\mathbf{P}$, and $\mathbf{P}'$ is obtained from $\mathbf{P}$ by adding a new candidate $y$ who does not beat $x$ in a head-to-head majority comparison, then $x$ still wins in $\mathbf{P}'$. 
\item \textit{independence of clones}: see Section \ref{ClonesSection}.
\end{itemize}
\end{itemize}

Several of these axioms involved formalizing auxiliary relations or operations on profiles. For example, monotonicity requires the ternary relation of one profile being related to another by a \textit{simple lift} of a candidate $x$, meaning that $x$ may go up in voters' rankings but the rest of their rankings remains the same:

\begin{itemize}
\item[] \texttt{\textcolor{blue}{def} \textcolor{orange}{simple\_lift} $\{$V X : \textcolor{blue}{Type}$\}$ : Prof V X $\to$ Prof V X $\to$ X $\to$ Prop := 
    \textcolor{blue}{$\lambda$} P$'$ P x, ($\forall$ (a $\neq$ x) (b $\neq$ x) i, P i a b $\leftrightarrow$ P$'$ i a b) 
    $\wedge$  \\$\forall$ a i, ((P i x a $\to$ P$'$ i x a) $\wedge$ (P$'$ i a x $\to$ P i a x))}
 \end{itemize}
The variable-voter and variable-candidate axioms involve moving from one profile to another with a different set of voters or candidates. In fact, it is most convenient to formalize these axioms using the operations of \textit{removing} a voter or candidate from a profile, moving from a profile of type \texttt{Prof V X} to a profile of type  \texttt{ Prof $\{$v : V // v $\neq$ i$\}$ X}, in the case of removing a voter \texttt{i}, or of type \texttt{Prof V $\{$x : X // x $\neq$ b$\}$}, in the case of removing a candidate \texttt{b}. See Section~\ref{ClonesSection} for the definition of the \textcolor{orange}{\texttt{minus\_candidate}} operation.

We stress that the axiom definitions we formalized are the standard ones used throughout the literature, not ones tailored to any particular VSCC, so others can immediately use our formalizations to verify results for any VSCCs.

\section{Theorems}\label{Theorems}

As a proof of concept of verifying theorems using our formal framework, we verified most of the results concerning the Split Cycle voting method in \cite{HP2020b}. In particular, we verified the equivalence of two definitions of Split Cycle (one quantifying over all cycles containing $x$ and $y$, the other quantifying over only paths from $y$ to $x$) and that Split Cycle satisfies all of the axioms listed in Section~\ref{VotingAxioms}. However, we emphasize that Split Cycle was only chosen as a test case. Our formalization of the basic objects and axioms of voting theory can be used to verify properties of any other voting method. 

Moreover, verifying the properties of Split Cycle required formalizing a number of general-purpose lemmas that will be needed---and can now be used---to verify the properties of other voting methods. Indeed, one of the benefits of a formalization project such as ours is to extract those general-purpose lemmas, which may be buried in proofs for a particular voting method, so that future formalization efforts can move more quickly to formalizing sophisticated results.

\subsection{Graph-theoretic background}

Before formalizing voting-theoretic proofs, we had to build up basic infrastructure for reasoning about cycles, walks, and paths in graphs, such as rotating and reversing cycles and converting walks to paths, which was not available in Mathlib. For example, to convert walks to paths, we defined an inductive type\footnote{We can eliminate `\texttt{noncomputable}' if we assume that equality for \texttt{X} is decidable, adding `\texttt{[decidable\_eq X]}' as an implicit argument. }:
\begin{itemize}
\item[] \texttt{\textcolor{magenta}{noncomputable} \textcolor{blue}{def} \textcolor{orange}{to\_path} $\{$X : \textcolor{blue}{Type}$\}$ : list X $\to$ list X}
\item[] \texttt{| [] := []}
\item[] \texttt{| (u :: p) := \textcolor{blue}{let} p$'$ := to\_path p \textcolor{blue}{in}}
\item[]    \quad\texttt{\textcolor{magenta}{if} u $\in$ p$'$ \textcolor{magenta}{then} (p$'$.drop (p$'$.index\_of u)) \textcolor{magenta}{else} (u :: p$'$)}
 \end{itemize}
Hence \texttt{to\_path} maps the empty list to itself, and given a list \texttt{u :: p} constructed by adding \texttt{u} to the front of the list \texttt{p},  if \texttt{u} is an element of \texttt{to\_path~p}, we output the result of dropping from \texttt{to\_path p} all elements before \texttt{u} in the list, and otherwise we add \texttt{u} to the front of \texttt{to\_path p}. A significant part of the formalization effort was proving needed properties of  \texttt{to\_path} and other operations on lists.

\subsection{Reasoning about margins}

In addition to developing graph-theoretic infrastructure, we formalized a number of general-purpose lemmas for reasoning about majority margins between candidates and the relation between changes in profiles and changes in margins. These lemmas are applicable to all voting methods for which the selection of winners is invariant between profiles with the same majority margins matrices---so-called C2 voting methods \cite{Fishburn1977} (e.g., Minimax, Ranked Pairs, Beat Path, Split Cycle, and even Borda). To take one example, when reasoning about monotonicity (recall Section \ref{VotingAxioms}) for C2 methods, a key fact is that a simple lift of a candidate $x$ cannot  increase the margin of any candidate over~$x$:
\begin{itemize}
\item[] \texttt{\textcolor{blue}{lemma} \textcolor{orange}{margin\_lt\_margin\_of\_lift}  $\{$V X : \textcolor{blue}{Type}$\}$ (P P$'$ : Prof V X)} \\  \texttt{[fintype V] (y x : X) :} \\
\texttt{simple\_lift P$'$ P x $\to$ margin P$'$ y x $\leq$ margin P y x := $\dots$}
\end{itemize}
Other general-purpose lemmas about margins that we formalized concern how margins change when adding a voter who ranks $x$ first---for verifying the axiom of positive involvement---or last---for verifying the axiom of negative involvement---or how clones of a candidate all bear the same margins to a given non-clone---for verifying the axiom of independence of clones. These are basic lemmas that anyone wishing to verify the relevant properties of C2 voting methods will need. 

For some of the most elementary facts, Lean's automation was adequate to complete proofs without our help. For example, the fact that reversing all voters rankings reverses all margins---used in verifying the axiom of reversal symmetry---was provable using Lean's \texttt{obviously} tactic:

\begin{itemize}
\item[] \texttt{\textcolor{blue}{def} \textcolor{orange}{reverse\_profile} $\{$V X : \textcolor{blue}{Type}$\}$ : Prof V X $\to$ Prof V X := \\ $\lambda$ P v x y, P v y x} \\
\item[]  \texttt{\textcolor{blue}{lemma} \textcolor{orange}{margin\_reverse\_eq}  $\{$V X : \textcolor{blue}{Type}$\}$ [fintype V] (P : Prof V X)  \\}
\texttt{(a b : X) : margin (reverse\_profile P) b a = margin P a b :=}

\texttt{\textcolor{blue}{begin}}

  \quad \texttt{obviously,}
  
\texttt{\textcolor{blue}{end}}
\end{itemize}
Similarly, Lean's \texttt{obviously} tactic immediately proves that removing a candidate does not change the margins between remaining candidates. By contrast, proofs of lemmas like \texttt{\textcolor{orange}{margin\_lt\_margin\_of\_lift}} required more human input.

\subsection{Example: independence of clones}\label{ClonesSection}

Our most involved formalization was of the proof from \cite{HP2020b} that Split Cycle satisfies Tideman's \cite{Tideman1987} axiom of independence of clones. A set $C$ of two or more candidates is a set of \textit{clones} in a profile $\mathbf{P}$ if no voter ranks any candidates outside of $C$ in between two candidates from $C$. Given a particular candidate $c$, we say that a nonempty set $D$ of candidates (not containing $c$) is a set of \textit{clones of $c$} if $D\cup \{c\}$ is a set of clones. In Lean, we formalize this as follows:

\begin{itemize}
\item[] \texttt{\textcolor{blue}{def} \textcolor{orange}{clones} $\{$V X : \textcolor{blue}{Type}$\}$ (P : Prof V X) (c : X)}

\texttt{(D : set $\{$x : X // x $\neq$ c$\}$) : Prop := }

\texttt{D.nonempty $\wedge$ ($\forall$ (c$'$ $\in$ D) (x : $\{$x : X // x $\neq$ c$\}$) (i : V),}

\texttt{x $\not\in$ D $\to$ ((P i c x $\leftrightarrow$ P i c$'$ x) $\wedge$ (P i x c $\leftrightarrow$ P i x c$'$)))}
\end{itemize}
Independence of clones for VSCCs states that (i) removing a clone from a profile should not change which non-clones win and (ii) removing a clone from a profile should not change whether at least one clone is among the winners (though which clone wins is allowed to change upon removing a clone). To formalize this, we need a way of removing a candidate from a profile, accomplished as follows:
\begin{itemize}
\item[] \texttt{\textcolor{blue}{def} \textcolor{orange}{minus\_candidate} $\{$V X : \textcolor{blue}{Type}$\}$ (P : Prof V X) (b : X) :} \\
\texttt{ Prof V $\{$x : X // x $\neq$ b$\}$ := \textcolor{blue}{$\lambda$} v x y, P v x y}
\end{itemize}
Thus, \texttt{minus\_candidate} takes in a profile \texttt{P} for \texttt{V} and \texttt{X}, as well as a candidate \texttt{b} from \texttt{X}, and outputs the profile for \texttt{V} and \texttt{$\{$x : X // x $\neq$ b$\}$} that agrees with \texttt{P} on how every voter ranks the candidates other than \texttt{b}. Using \texttt{minus\_candidate}, we formalize condition (i) of independence of clones as follows:
\begin{itemize}
\item[] \texttt{\textcolor{blue}{def}   \textcolor{orange}{non\_clone\_choice\_ind\_clones} $\{$V X : \textcolor{blue}{Type}$\}$ (P : Prof V X)}

\texttt{(c : X) (D : set $\{$x : X // x $\neq$ c$\}$) : VSCC $\to$ Prop := \textcolor{blue}{$\lambda$} F, }

\texttt{clones P c D $\to$ ($\forall$  a : $\{$x : X // x $\neq$  c$\}$, a $\not\in$ D $\to$ }

\texttt{(a.val $\in$ (F V X P) $\leftrightarrow$ a $\in$ (F V $\{$x : X // x $\neq$ c$\}$} 

\texttt{(minus\_candidate P c))))}
\end{itemize}
Since \texttt{$\{$x : X // x $\neq$  c$\}$} is a subtype of \texttt{X},  \texttt{a} consists of an element of \texttt{X}, called \texttt{a.val}, together with a proof that \texttt{a.val $\neq$ c}. Since \texttt{F V X P} is a set of elements of \texttt{X}, we must write `\texttt{a.val $\in$ (F V X P)}' instead of `\texttt{a $\in$ (F V X P)}'. Finally, we can state that Split Cycle satisfies part (i) of independence of clones as follows:

\begin{itemize}
\item[] \texttt{\textcolor{blue}{theorem} \textcolor{orange}{non\_clone\_choice\_ind\_clones\_split\_cycle} $\{$V X : \textcolor{blue}{Type}$\}$}

\texttt{[fintype V] (P : Prof V X) (c : X) (D : set $\{$x : X // x $\neq$ c$\}$) : }
\texttt{non\_clone\_choice\_ind\_clones P c D split\_cycle := ...}
\end{itemize}
The formalization of part (ii) of independence of clones is similar. The proof that Split Cycle satisfies independence of clones involves manipulating paths in the majority graph of a profile---in particular, replacing all clones in a path by a distinguished clone and then eliminating repetitions of candidates in the resulting sequence using the \texttt{to\_path} operation.

\section{Conclusion}\label{Conclusion}

Our goal was to set up a general framework for formally verifying results in voting theory. One of the benefits of such a formalization project, beyond certifying the correctness of results, is that it forces formalizers to think about how the fundamental notions of the field ought to be formulated (see, e.g., our definitions of VSCCs and VCCRs). We expect such benefits to accrue for formalization projects in other areas of mathematical social science or even natural science.

What did we learn from the verification stage of our project? As usual in formalization, we caught some omitted assumptions (e.g., of nonemptiness) in definitions needed to prove results about the Split Cycle voting method in a draft of \cite{HP2020b}, prompting corrections. A more striking lesson of formalizing these results is how little depends on assumptions about properties of voter preferences. While it was initially assumed in \cite{HP2020b} that voter preference relations are linear orders,\footnote{This assumption was made for its simplifying consequence of rendering the majority margin between two candidates the canonical measure of the strength of majority preference between candidates. Cf.~the following paragraph in the main text.} the full strength of this assumption turned out not to be used in any proofs we formalized. In fact, most results work with no assumptions about voter preferences at all (except the default asymmetry of strict preference). The only exception was the Pareto principle, whose proof used the acyclicity of voter preferences. It would be fascinating to see exactly what properties of voter preferences are needed in formalized proofs of properties of other voting methods.

It would also be desirable to abstract away from the definition of margin in Definition \ref{MarginDef} to define voting methods and prove theorems in terms of an abstract relation $(a,b)\succ_\mathbf{P} (c,d)$ expressing that the strength of majority preference for $a$ over $b$ in profile $\mathbf{P}$ is stronger than the strength of majority preference for $c$ over $d$ in $\mathbf{P}$. One definition of $(a,b)\succ_\mathbf{P} (c,d)$ is that $Margin_\mathbf{P}(a,b)>Margin_\mathbf{P}(c,d)$. But if voter preference relations are non-linear, there are alternative, inequivalent definitions of $(a,b)\succ_\mathbf{P} (c,d)$, such as the \textit{winning votes} definition: $\big|\{i\in V(\mathbf{P})\mid a\mathbf{P}_ib\}\big| \geq \big|\{i\in V(\mathbf{P})\mid c\mathbf{P}_id\}\big|$. Schulze \cite{Schulze2011} considers other definitions besides margins and winning votes, and rather than settling for any one of them, lays down general axioms on the relation of relative strength of majority preference needed for proofs to go through. This is a natural next step in our project, which formalization facilitates: identify which of the properties of the $\succ_\mathbf{P}$ relation are actually used in our formalized proofs and then assume only those properties.

Another natural next step would be to extend the verification of voting axioms beyond voting methods, such as Split Cycle, that are based on head-to-head majority comparisons to voting methods, such as Instant Runoff, that are based on iterative elimination procedures. Instant Runoff can be defined \textit{recursively}: a candidate $x$ is an Instant Runoff winner in a profile $\mathbf{P}$ if either $x$ is the only candidate in $\mathbf{P}$ or there is some candidate $y$ who receives the fewest first-place votes in $\mathbf{P}$, and $x$ is an Instant Runoff winner in $\mathbf{P}_{-y}$.\footnote{This is the ``parallel universe'' version of Instant Runoff. An alternative version \cite[p.~7]{Taylor2008} states that the Instant Runoff winners in $\mathbf{P}$ are the Instant Runoff winners in the profile $\mathbf{P}_{-Y}$ obtained from $\mathbf{P}$ by removing the set $Y$ of \textit{all} candidates who receive the fewest first-place votes in $\mathbf{P}$, if $Y\subsetneq X(\mathbf{P})$; otherwise all candidates~win. For another example of a recursively-defined voting method, see \cite{HPSV2021}.} As Lean offer natural ways of defining functions recursively and writing proofs by induction \cite[\S~8]{Avigad2021}, we expect Lean to be well suited to verifying properties of recursively-definable voting methods such as Instant Runoff.

With its axiomatic approach and discrete mathematical character, voting theory is especially amenable to formal verification. Moreover, given the importance of democratic decision making in society, we find it desirable to formally verify that democratic decision procedures have the desirable properties claimed for them. We have done so for one recently proposed voting method, but we would like to see this done for all methods proposed for use in democratic elections.

\subsection*{Acknowledgement}

We thank the Lean Zulip chat community, the Berkeley Lean Seminar, and Jeremy Avigad for advice about using Lean and the two anonymous referees for helpful feedback on our paper.

 \bibliographystyle{splncs04}
 \bibliography{VotingTheoryLean}
\end{document}